\documentclass[journal,twoside,web]{ieeecolor}

\usepackage{generic}
\usepackage{cite}
\usepackage{amsmath,amssymb,amsfonts}
\usepackage{algorithmic}
\usepackage{graphicx}
\usepackage{algorithm,algorithmic}
\usepackage{hyperref}
\hypersetup{hidelinks=true}
\usepackage{breqn}
\usepackage{textcomp}

\def\BibTeX{{\rm B\kern-.05em{\sc i\kern-.025em b}\kern-.08em
    T\kern-.1667em\lower.7ex\hbox{E}\kern-.125emX}}
\begin{document}
\title{GuidedMorph: Two-Stage Deformable Registration for Breast MRI}
\author{Yaqian Chen, Hanxue Gu, Haoyu Dong, Qihang Li, Yuwen Chen, Nicholas Konz, Lin Li, Maciej A. Mazurowski
\thanks{Yaqian Chen, Hanxue Gu, Haoyu Dong, Yuwen Chen, and Nicholas Konz are with the Department of Electrical and Computer Engineering, Duke University, Durham, NC 27703, USA}
\thanks{Qihang Li is with the Department of Biostatistics \& Bioinformatics, Duke University, Durham, NC 27703, USA}
\thanks{Lin Li was with the Department of Biostatistics \& Bioinformatics, Duke University, Durham, NC 27703, USA. Now, Lin Li is with the Department of Health Outcomes and Biomedical Informatics, College of Medicine, University of Florida, Gainesville, FL 32611}
\thanks{Maciej A. Mazurowski is with the Departments of Electrical and Computer Engineering, Radiology, Biostatistics \& Bioinformatics, and Computer Science, Duke University, Durham, NC 27703, USA}}

\maketitle

\begin{abstract}
Accurately registering breast MR images from different time points enables the alignment of anatomical structures and tracking of tumor progression, supporting more effective breast cancer detection, diagnosis, and treatment planning. However, the complexity of dense tissue and its highly non-rigid nature pose challenges for conventional registration methods, which primarily focus on aligning general structures while overlooking intricate internal details. To address this, we propose \textbf{GuidedMorph}, a novel two-stage registration framework designed to better align dense tissue. In addition to a single-scale network for global structure alignment, we introduce a framework that utilizes dense tissue information to track breast movement. The learned transformation fields are fused by introducing the Dual Spatial Transformer Network (DSTN), improving overall alignment accuracy. A novel warping method based on the Euclidean distance transform (EDT) is also proposed to accurately warp the registered dense tissue and breast masks, preserving fine structural details during deformation. The framework supports paradigms that require external segmentation models and with image data only. It also operates effectively with the VoxelMorph and TransMorph backbones, offering a versatile solution for breast registration. We validate our method on ISPY2 and internal dataset, demonstrating superior performance in dense tissue, overall breast alignment, and breast structural similarity index measure (SSIM), with notable improvements by over 13.01\% in dense tissue Dice, 3.13\% in breast Dice, and 1.21\% in breast SSIM compared to the best learning-based baseline.
\end{abstract}

\begin{IEEEkeywords}
Image registration, Deep learning, Breast MRI
\end{IEEEkeywords}

\section{Introduction}
In recent decades, breast MRI has become a critical tool in the early detection of breast cancer \cite{intro1}, particularly for high-risk populations \cite{intro2}.
Its superior sensitivity and level of detail compared to traditional imaging methods such as mammography and ultrasound enable it to capture sufficient information to identify early-stage cancers that other modalities might miss \cite{intro_sen_1,intro_sen_2,intro_sen_3,intro_sen_4}.
While breast MRI excels in sensitivity for early cancer detection, it suffers from high false-positive rates, particularly in initial screenings, necessitating several subsequent scans for accurate assessment \cite{heller2019mri}.  

Thus, during diagnostic assessments, it is common for radiologists to track certain changes in the breast over time through sequential scans. In this context, dense tissue alignment is particularly important due to its clinical significance \cite{denseimpo1,denseimpo2,denseimpo3}; rapid changes or growth in these regions are often associated with a heightened risk of developing breast cancer \cite{denseimpo4,maloney2022breast}. Consequently, accurately registering dense tissue across MRI scans is essential for capturing subtle transformations that may signal the onset or progression of malignancy.

\begin{figure}[htbp]
  \centering
  \includegraphics[width=0.45\linewidth]{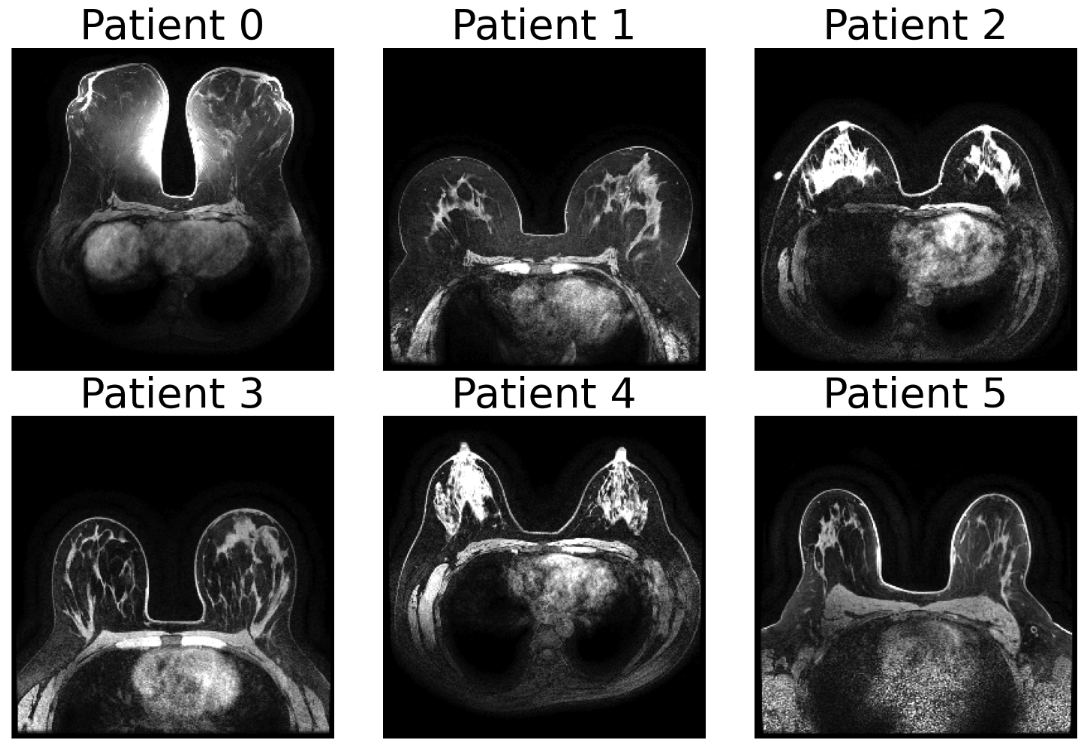}
  \hspace{0.04\linewidth}
  \includegraphics[width=0.45\linewidth]{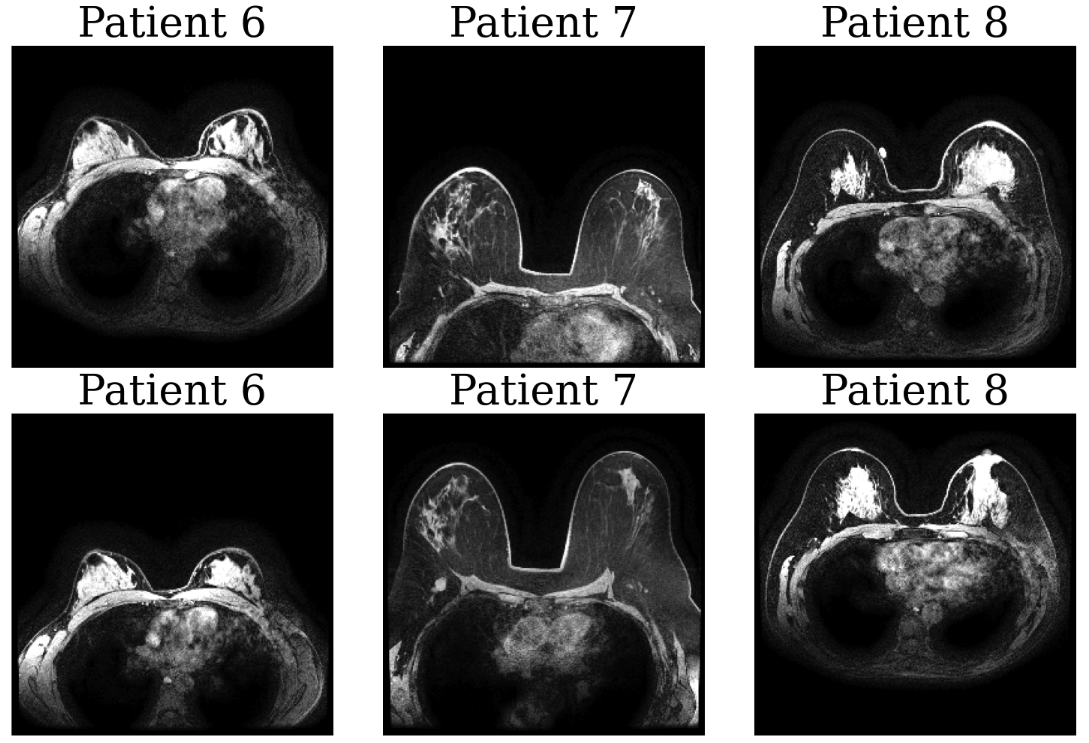}
  \caption{\textbf{Inter-patient and intra-patient variability in breast MRI scans.} Middle slices from different MRI volumes illustrate differences across patients (\textit{left}) and within the same patient (\textit{right}).}
  \label{fig:difference}
\end{figure}

Despite the wide array of medical image registration techniques currently available, achieving accurate breast MRI registration—especially the precise alignment of dense tissue—continues to be a significant challenge \cite{french2023diffeomorphic,ham2023improvement}. This difficulty stems largely from the complexity of dense tissue, which can vary considerably (as shown in Fig. \ref{fig:difference}), and the high non-rigidity of the breast (as opposed to other commonly registered modalities like brain MRI \cite{mehrabian2018deformable,heinrich2022voxelmorph++}).



In this paper, we propose a two-stage registration framework, \textbf{GuidedMorph}, that is particularly designed for the challenging tasks such as breast MRI registration. In additional to the structure of VoxelMorph \cite{voxelmorph} and its variants \cite{chen2022transmorph,kim2021cyclemorph}, our method introduces a dedicated step to improve alignment in information-rich regions, such as dense tissue. It also integrates a publicly available VNet-based segmentation model for dense tissue and breast boundaries \cite{vnet, chen2025breast}, using pre-trained weights that require no fine-tuning or extra annotations. To further simplify the pipeline, we propose a non-learning \textbf{Dense Tissue Extraction} strategy that identifies complex dense tissue regions based on Frangi vesselness filte. Furthermore, with the observation that traditional warping methods tend to lose fine structural details (See \ref{fig:EDT_result}) when applied to binary masks, especially for sparse or anatomically complex regions such as dense tissue, we propose a more robust alternative. A novel warping method based on the Euclidean distance transform (EDT) is proposed, which converts binary masks into continuous-valued distance maps prior to deformation. The novel transformation method consistently demonstrates performance improvements by preserving structural integrity.

The main contributions of this work are summarized as follows:
\begin{itemize}
    \item We propose GuidedMorph, a novel dual-scale alignment architecture, combining a global structure alignment network with a dense tissue tracking module that utilizes the Dual Spatial Transformer Network to improve the fidelity of non-rigid dense tissue.
    \item This registration pipeline is flexible, supporting both scenarios where external segmentation masks are available and cases where only image data is provided.
    \item A new detail-preserving mask warping algorithm is proposed, utilizing the EDT to preserve fine structural details during mask deformation. This approach is significantly effective in preserving thin and fine dense tissue structures, leading to a 2.44\% improvement in performance compared to conventional mask warping.
    \item Compared to the best learning-based baseline, our method achieves the state-of-the-art, improving the performance by over 13.01\% in dense tissue Dice, 3.13\% in breast Dice, and 1.21\% in breast SSIM on both internal and external datasets.
\end{itemize}

We will release our code upon paper acceptance.

\section{Related Work}\label{sec:relatedWorks}
\subsection{Medical Image Registration}
Medical image registration has increasingly attracted attention due to its essential role in aligning anatomical structures across different imaging modalities, time points, or patients, enabling tasks such as longitudinal disease tracking, atlas construction, and image-guided interventions \cite{imran2010medical, royenhancing, wyawahare2009image}. Traditional registration approaches are broadly categorized into rigid, affine, and deformable methods \cite{maintz1998survey}. Rigid and affine registrations, which are simpler in nature, are well-supported by numerous existing tools, such as Automated Image Registration (AIR) \cite{woods1998automated}, Correspondence of Closest Gradient Voxels (COCGV) \cite{ostuni1997correspondence}, and Advanced Normalization Tools (ANTS) \cite{avants2009advanced}. However, the assumption of rigidity can be limiting for realistic medical image registration, particularly for breast tissue. To mitigate this limitation, recent medical image registration research has shifted toward deformable registration algorithms \cite{sotiras2013deformable, mansilla2020learning}, which model non-linear anatomical variability and enable finer alignment, particularly in soft-tissue applications such as breast MRI \cite{voxelmorph, chen2022transmorph, kim2021cyclemorph}. Early deformable methods relied on optimization-based strategies \cite{avants2008symmetric, ourselin2001reconstructing}, conducting optimization iteratively for every registration pair. In order to shorten the inference time, recent studies such as VoxelMorph \cite{voxelmorph} and TransMorph \cite{chen2022transmorph} have demonstrated promising results by accelerating registration through unsupervised training paradigms.  

\subsection{Non-Deep Learning Medical Image Registration Methods}
Traditional non-deep learning-based deformable registration algorithms typically optimize a similarity metric between fixed and moving images through iterative transformation updates via three main components: a deformation model, an objective function, and an optimization method \cite{sotiras2013deformable}. The deformation model defines the type of transformation applied to $m$, ranging from physical models such as elastic body models \cite{broit1981optimal,bajcsy1989multiresolution,gee1998elastic,rabbitt1995mapping,pennec2005riemannian,droske2004variational}, viscous fluid flow models \cite{christensen1996deformable,wang2000physical,d2003viscous,chiang2008fluid}, and diffusion models \cite{thirion1998image,pennec1999understanding,vercauteren2007insight,vercauteren2007non} to models based on interpolation theory \cite{bookstein1989principal,bookstein1991thin,rohr2001landmark,rueckert2006diffeomorphic}. The objective function typically includes similarity metrics such as Cross-Correlation (CC) \cite{bourke1996cross}, Mean Squared Error (MSE) \cite{fisher1920012}, and Mutual Information (MI) \cite{kraskov2004estimating}, as well as a regularization term used to balance the accuracy of the transformation with the smoothness of the deformation field. The optimization strategy varies depending on the number of degrees of freedom of the deformation model, which can range from six for global rigid transformations to millions for non-parametric dense transformations. The increased dimensionality of the state space enriches the descriptive power of the model, but also necessitates more complex and time-consuming optimization algorithms \cite{sotiras2013deformable}. This trade-off between model complexity and computational efficiency is a key consideration in non-learning-based registration approaches \cite{yang2022ransacs}, and as we will show in the next section, the high model expressivity yet scalable training of deep learning makes it particularly suited for realistic registration problems.

\subsection{Deep Learning-based Medical Image Registration Methods}
To address the aforementioned complexity-efficiency tradeoff of non-learning-based registration approaches, many deep learning-based registration methods have been proposed \cite{liu2019multimodal, liu2019image, miao2018dilated, miao2016cnn}.
Different from traditional methods that focus on iteratively minimizing a predefined function eq. \eqref{reg_func}, deep neural network (DNN)-based methods are designed to learn the optimal registration field given a training dataset. This approach effectively learns a comprehensive global representation of image registration, facilitating the alignment of previously unseen pairs of volumes and significantly reducing registration time via the single-forward-pass nature of neural networks.

In 2019, Balakrishnan et al. introduced a registration framework called VoxelMorph, which combined convolutional neural networks (CNNs) with spatial transformer networks (STN) \cite{voxelmorph}. This framework integrates the image feature extraction capabilities of CNNs with the spatial transformation abilities of STNs, enabling robust and efficient alignment of image pairs. Compared to traditional registration algorithms, this method offered significantly higher flexibility and drastically shortened registration time. However, the default single-scale U-Net architectures that utilize a direct spatial transformer loss often perform poorly for registration outside of the cranial area, falling short of achieving state-of-the-art results in registration tasks with large displacement \cite{heinrich2022voxelmorph++}.

To overcome these limitations, some other approaches have implemented multilevel schemes that effectively handle both small/local and large/global deformations \cite{bajcsy1989multiresolution, de2019deep, hering2019mlvirnet}. However, many multilevel registration networks merely downsample image features without directing the network's focus to the region of interest \cite{hering2021cnn}. This is especially crucial for breast registration, where dense tissue occupies a small portion of the volume. Therefore, in this paper, we propose a two-stage VoxelMorph-based registration method that balances global alignment with the registration of smaller, complex yet crucial regions. Note that crucial regions can either be defined by the user via mask input or automatically extracted based on frangi vesselness filter \cite{frangi1998multiscale}, ensuring a balanced and accurate registration process shown in Sec. \ref{sec:Method}.

\subsection{Registration Methods for Breast MRI}
Breast MRI registration is a challenging task and has received limited attention in recent literature. Early studies, such as Zuo et al. (1996) \cite{zuo1996automatic}, introduced breast MRI registration based on rigid transformations, acknowledging the need to eliminate patient movement-induced misalignment and distortions in imaging results. Subsequent research has focused on refining and extending these methods to improve registration accuracy. Commonly used techniques are based on biomechanical modeling \cite{krishnan1999linear, eiben2016surface, lee2008biomechanical}. These methods utilize finite element methods (FEMs) for solving differential equations to predict the displacement of tissue within the breast. However, these models are limited in their ability to accurately capture and correct the large deformations that the breast can undergo in practice \cite{french2023diffeomorphic}, and it is computationally complicated to develop a biomechanical model for each patient individually.

Some deep-learning-based registration models have also been utilized to address the challenge of breast MRI registration \cite{french2023diffeomorphic, zhang2020unsupervised, ying2022two}. However, these algorithms primarily evaluate their performance based on tumor detection, often neglecting the accuracy of registration in normal breast MRI cases and rarely comparing their effectiveness against existing registration methods in terms of entire-volume similarities and overall registration quality.

\section{Method}\label{sec:Method}
This work proposes a novel network framework that demonstrates superior performance on breast MRI deformable registration. The following sections present (a) the problem definition (Sec. \ref{sec:problem definition}), (b) an overview of the GuidedMorph framework (Sec. \ref{sec:methods:overview}), and additional methodological details.

\subsection{Problem Definition} \label{sec:problem definition}
Given a pair of breast MRI scans, our registration method aims to align a moving (or \textit{source}) volume to another fixed (\textit{target}) volume. Formally, this is defined by the optimization problem of finding 
\begin{equation}\label{reg_func}
    \hat{\phi} = \underset{\phi}{\operatorname{argmin}}\,E_{sim}(m\circ\phi, f) + \lambda R(\phi)
\end{equation}
where $m$ and $f$ represent the moving volumes and fixed volumes, respectively, $\phi$ is the \textit{deformation field} used to deform $m$ to align with $f$ via the composition $m \circ \phi$. $E_{sim}(m \circ \phi, f)$ measures the similarity between fixed and warped moving volumes, the commonly used measurements include Cross-Correlation (CC) \cite{bourke1996cross} and Mean Squared Error (MSE) \cite{fisher1920012}. Finally, $R(\phi)$ acts as a regularization function that enforces the smoothness of the deformation field, and the hyperparameter $\lambda$ governs the regularization trade-off between fidelity to the data with the smoothness of the transformation.

\subsection{GuidedMorph Overview}
\label{sec:methods:overview}
\begin{figure*}[htbp]
    \centering
    \includegraphics[width=1\textwidth]{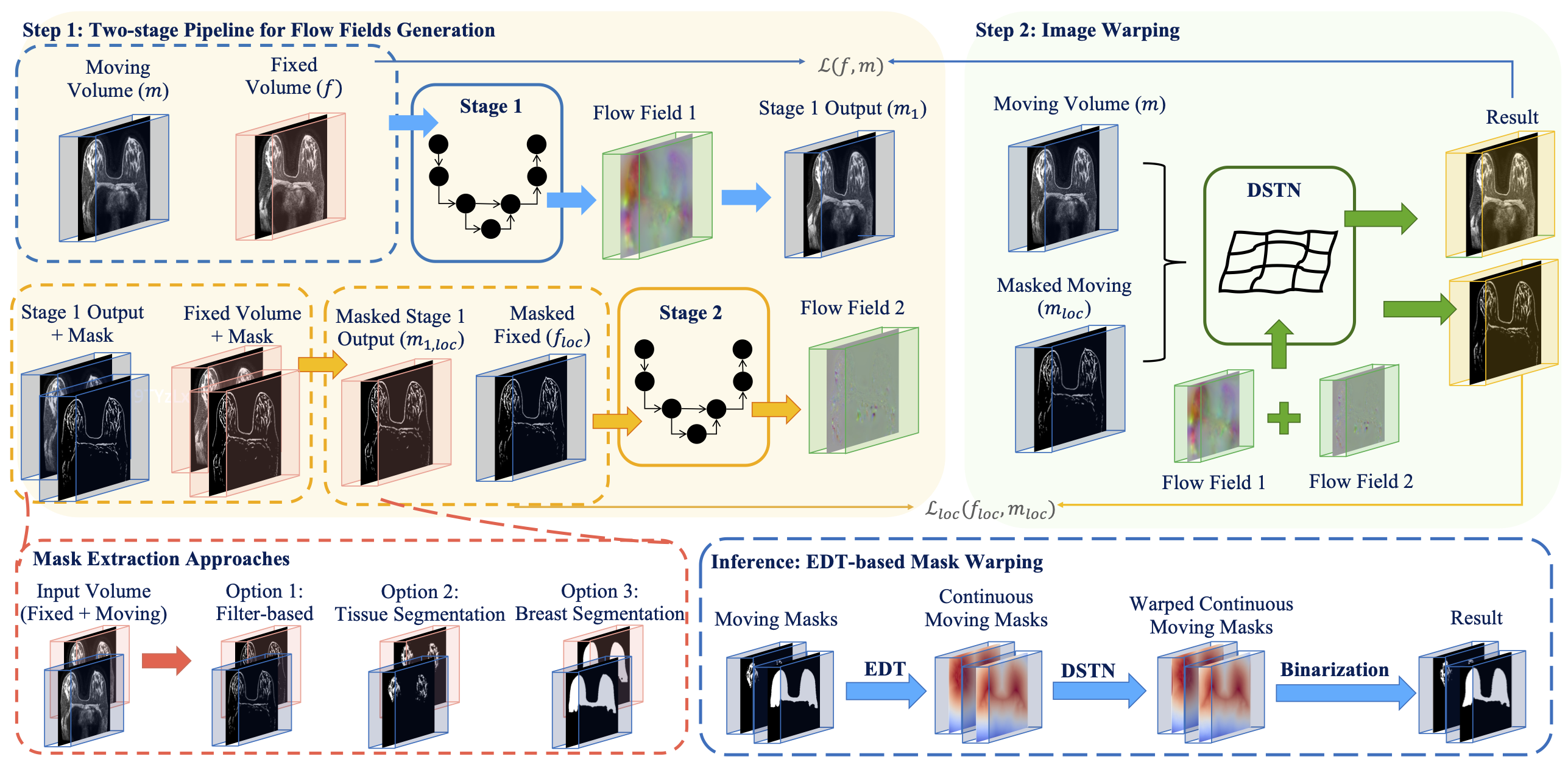}
    \caption{\textbf{GuidedMorph method overview:} the top branch shows Stage 1 (\textit{blue}), which aligns the moving and fixed volumes to estimate a global flow field. The bottom branch shows Stage 2 (\textit{yellow}), where extracted features—obtained via filter-based methods, dense tissue segmentation, or breast segmentation—are concatenated with the stage 1 output to guide localized refinement. The framework is supervised by both global image similarity loss and localized feature-based loss. EDT-based mask warping pipeline at the bottom right demonstrates our novel mask warping algorithm during inference stage.}
    \label{fig:pipleline}
\end{figure*}
Our proposed registration framework (summarized in Fig. \ref{fig:pipleline}) is a two-stage registration method, where the first stage performs global alignment of the entire MRI volume, while the second stage refines the alignment in critical regions of interest, specifically dense tissue. The framework can be built upon different convolutional networks, including U-Net with a single stage similar to VoxelMorph \cite{voxelmorph}, and Transformer whose single-stage structure resembles TransMorph \cite{chen2022transmorph}. However, in contrast to traditional single-scale network architectures that struggle to capture fine details in the volume, the newly introduced second stage in our GuidedMorph learns to output deformation fields for dense tissue and allows the motion from this fine but crucial region to guide the overall registration refinement.

The first stage of our pipeline takes a concatenated moving+fixed volume pair as its input. The fixed and moving volumes are two volumes belonging to the spatial domain $\Omega \subset \mathbb{R}^3$ , which are denoted as $m$ and $f$ respectively. A convolutional neural network (CNN) learns the displacement field $\mathbf{u}_{volume}$ for overall structure alignment between $m$ and $f$. The detailed architecture for CNN is described in the Implementation Sec. \ref{Implementation section}. The $\mathbf{u}_{volume}$ is a matrix defined over a 4-dimensional spatial domain. Specifically, when $m, f\in \mathbb{R}^{W\times H \times D}$ (where $W$, $H$, and $D$ stand for width, height, and depth of the volume, respectively), then the $\mathbf{u}_{volume}\in \mathbb{R}^{3\times W\times H \times D}$. The three channels of $\mathbf{u}_{volume}$ store the displacement along the $x, y, z$ axes, respectively.

After aligning the overall structure in the first stage, our second stage receives a concatenated volume pair supplemented by their corresponding extracted masks; these masks provide additional position information for dense tissue in volumes. In our algorithm, we provide three ways for mask generation: by employing our dense tissue extraction strategy based on Frangi vesselness filter shown in Sec. \ref{sec:methods:selection strategy}; through the learning-based dense tissue segmentation; or through the learning-based breast segmentation algorithms detailed in Sec. \ref{segmentation}. Masks generated from learning-based dense tissue segmentation provide precise position information for dense tissue, while breast mask and filter-extracted masks give less detailed information; however, filter-based extraction requires no external segmentation model, and breast masks are easy to annotate. The CNN in this stage learns the displacement field between the masked crucial regions in the fixed volume ($f_{loc}$) and stage 1 output moving volume ($m_{1, loc}$). The remaining voxels in the volume are then warped according to the movement of these crucial regions, ensuring that the entire volume is harmoniously transformed. The output displacement field from this stage, denoted as $\mathbf{u}_{mask}$, shares the same shape as $\mathbf{u}_{volume}$.

The deformation fields $\phi$ for GuidedMorph consist of two parts: $\phi_{volume} = Id + \mathbf{u}_{volume}$ and $\phi_{mask} = Id + \mathbf{u}_{mask}$,  where $Id$ is the identity transform. DSTN, which will be described in detail in Sec. \ref{sec:methods:twostageSTN}, is responsible for fusing the $\phi_{volume}$ and $\phi_{mask}$ into a single $\phi$, and further apply the spatial transformation to $m$ and $m_{loc}$ respectively. Notably, in the inference stage for binary mask warping, we introduce a EDT prior to warping to preserve structural fidelity during deformation; details are provided in Sec. \ref{sec:methods:distance transformation}. 

Stochastic gradient descent is utilized to optimize the respective parameters $\theta_1$ and $\theta_2$ for both CNN models simultaneously by minimizing a loss function shown in Eq. \eqref{loss_general}. The detailed implementation for the loss function is shown in Sec. \ref{sec:methods:loss}. We also employ dynamic loss weights to adjust the network focus over time. Particularly, the network initially concentrates on overall structure alignment given by stage 1 and then emphasizes dense tissue registration (stage 2). The detailed implementation for this dynamic focus is shown in Sec. \ref{Implementation section}.

\subsection{Dense Tissue Extraction Strategy}\label{sec:methods:selection strategy}
As detailed in Sec. GuidedMorph Overview \ref{sec:methods:overview}, we propose three approaches to extraction masks which offers dense tissue position information. The first approach utilizes our novel dense tissue extraction strategy, while the second and third approaches rely on learning-based breast and dense tissue segmentation model respectively. Although publicly available segmentation models and weights exist for breast and dense tissue \cite{lew2024publicly}, running inference with the provided code requires significant time and effort. To overcome this limitation, we introduce a novel dense tissue extraction strategy based on the Frangi vesselness filter \cite{frangi1998multiscale}. 

\begin{figure}[htbp]
    \centering
    \includegraphics[width=1\linewidth]{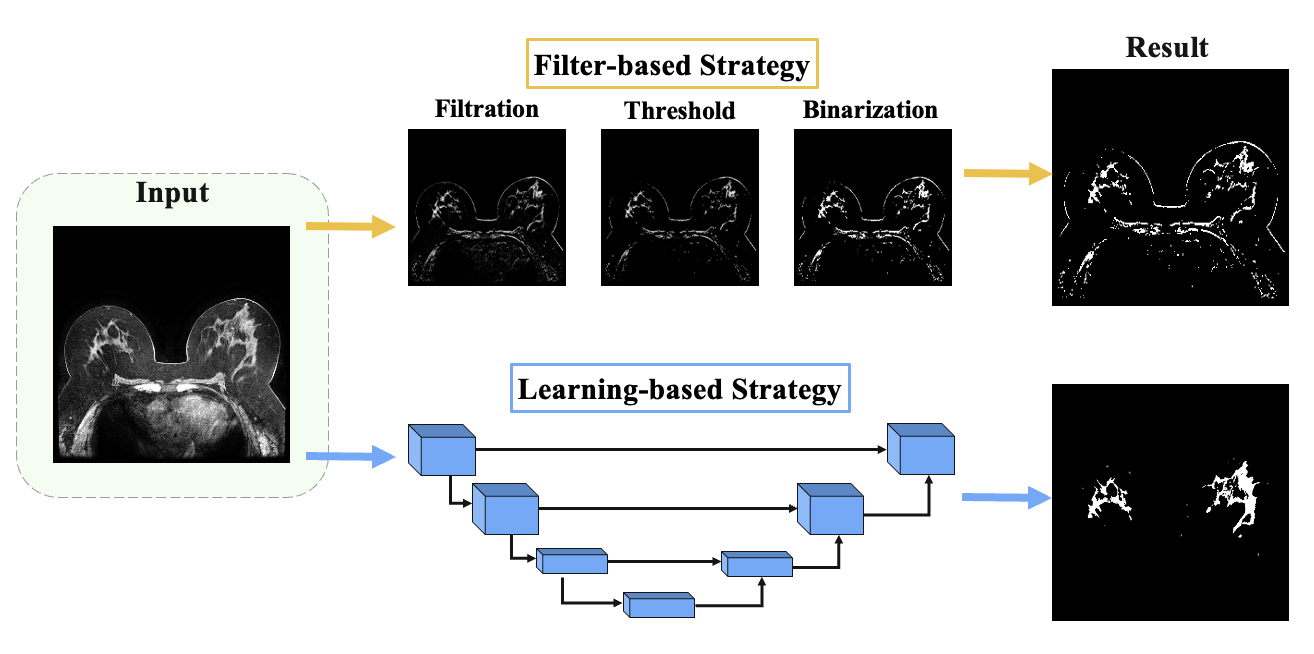}
    \caption{\textbf{Filter-based and learning-based breast dense tissue extraction overview and comparison:} The top branch (\textit{yellow}) shows the three steps of the filter-based strategy: Filtration, Thresholding, and Binarization. The bottom branch (\textit{blue}) shows the learning-based strategy. The outputs from the two strategies are compared in the Result column.}
    \label{fig:sub-volume selection pipeline}
\end{figure}

The pipeline for filter-based dense tissue extraction, along with its comparison to the segmentation model result, is illustrated in Fig. \ref{fig:sub-volume selection pipeline}. The extraction process contains three steps, specifically Filtration, Threshold, and Binarization. 

The filtration step employs the Frangi vesselness filter \cite{frangi1998multiscale} to extract tube-like structures in the volume. The Frangi vesselness filter, which is based on the Hessian matrix and its eigenvalues, is one of the mostly used methods to enhance the tube-like structures including blood vessels, neurites, and tissue. However, the Frangi vesselness filter retains noise in the chest, contours, as well as dense tissue when it is applied to breast MRI. 

\begin{figure}[htbp]
    \centering
    \includegraphics[width=1\linewidth]{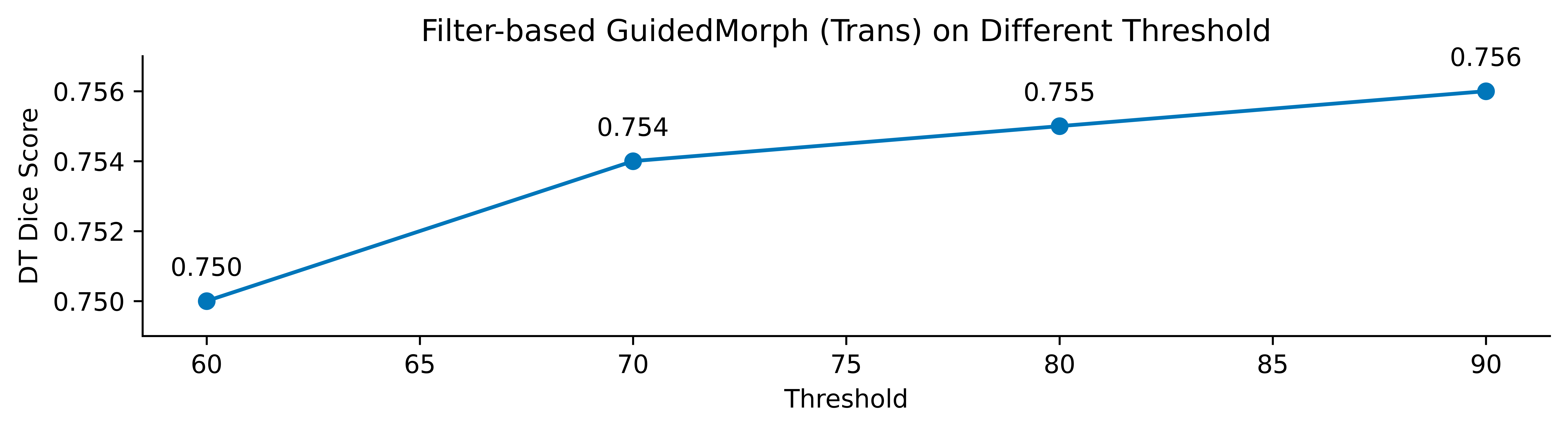}
    \caption{Quantitative evaluation of filter-based dense tissue extraction across different percentile thresholds.}
    \label{fig:threshold_result}
\end{figure}

To address this, we clip the low-intensity parts of the filtered volume in the threshold step to remove the noisy components in the chest. The 90th-percentile threshold is chosen based on both visual and empirical testing, as shown in Fig. \ref{fig:threshold} and Fig. \ref{fig:threshold_result}, respectively. The final binarization step binarizes the extracted structures, converting them into a usable binary mask to resemble the output for dense tissue segmentation model.

\begin{figure}[htbp]
    \centering
    \includegraphics[width=1\linewidth]{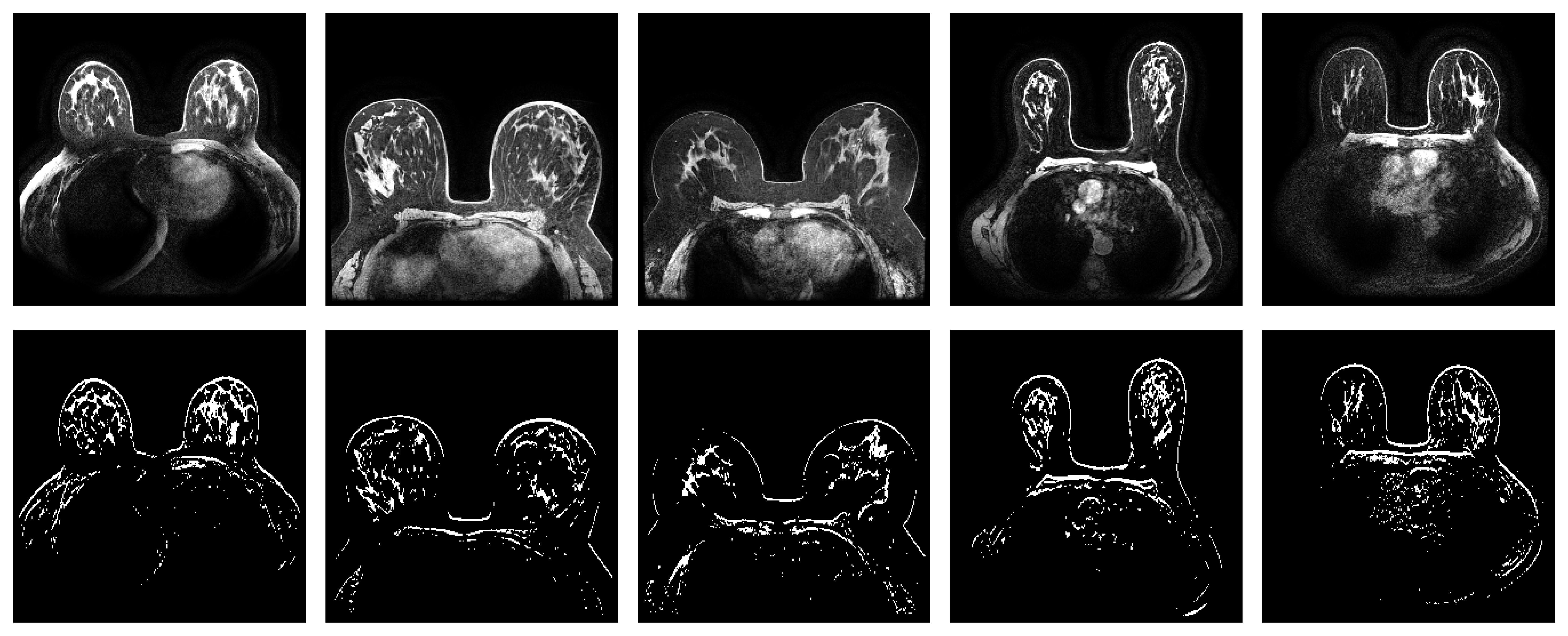}
    \caption{\textbf{Extracted masks based on the Frangi vesselness filter:} The top row shows the middle slices of the original volumes, and the bottom row shows the corresponding middle slices of the extracted masks.}
    \label{fig:filtered_volume}
\end{figure}

Although this thresholding process may affect some dense tissue details and some contour noise may still be retained, as shown in the result section of two braches in Fig. \ref{fig:sub-volume selection pipeline} and the dense tissue extraction results in Fig. \ref{fig:filtered_volume}, our experiments show that the rough dense tissue extraction based on filter still provides decent dense tissue information which improves the breast and dense tissue alignment significantly (Sec. \ref{result}). More detailed information on threshold selection is shown in Sec. \ref{sec:experiments:feature_extraction}.

\subsection{Dual Spatial Transformer Networks (DSTN)}
\label{sec:methods:twostageSTN}
The most intuitive way to apply both $\mathbf{u}_{volume}$ and $\mathbf{u}_{mask}$ to volumes such as $m$ and $m_{loc}$ is to apply the two transformations sequentially, each followed by interpolation. In practice, this is implemented by first applying $\mathbf{u}_{volume}$ via a spatial transformer network (STN), which includes both the transformation and the interpolation steps, to generate a half-warped volume denoted as $m_1$ (as shown in Eq. \eqref{Two-stage STN equation 1}).
\begin{equation}\label{Two-stage STN equation 1}
m_1= \sum_{\mathbf{q} \in \mathbb{Z}^3(\mathbf{p}')} m(\mathbf{q}) \prod_{d \in \{x,y,z\}} (1 - |\mathbf{p}'_d - \mathbf{q}_d|)
\end{equation}
Here, we adopt similar notation to \cite{voxelmorph}, where $\mathbf{p}$ represents each voxel ($\mathbf{p}\in \mathbb{R}^3$) inside the moving volume $m$, $\mathbf{p}' = \mathbf{p} + \mathbf{u}_{volume}\mathbf{(p)}$, $\mathbb{Z}^3(\mathbf{p})$ is the set of neighboring voxels of a given $\mathbf{p}$, and $d$ iterates over $\mathbb{R}^3$. 

Then, to generate the final warped volume ($m_2$), $m_1$ is put into the regular STN with $\mathbf{u}_{mask}$ (in Eq. \eqref{Two-stage STN equation 2}), where $\mathbf{p}_1$ is some voxel in $m_1$, and $\mathbf{p}'' = \mathbf{p}_1 + \mathbf{u}_{mask}(\mathbf{p}_1)$.

\begin{equation}\label{Two-stage STN equation 2}
 m_2 = \sum_{\mathbf{q} \in \mathbb{Z}^3(\mathbf{p}'')} m(\mathbf{q}) \prod_{d \in \{x,y,z\}} (1 - |\mathbf{p}''_d - \mathbf{q}_d|)
\end{equation}

\begin{figure}[htbp]
    \centering
    \includegraphics[width=1\linewidth]{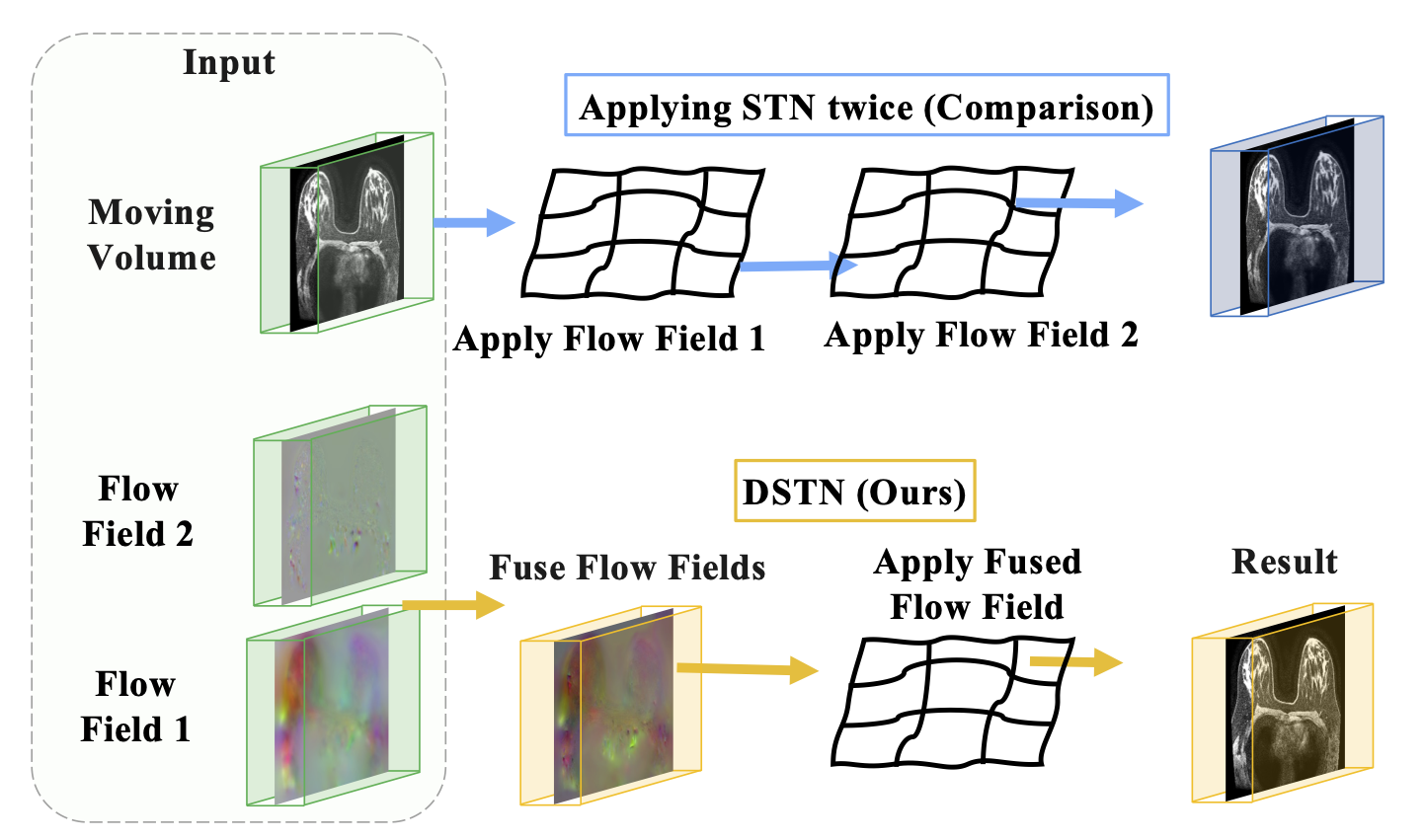}
    \caption{\textbf{Pipeline comparison between DSTN and applying STN twice:} As illustrated in top branch (\textit{blue}), the two-round STN approach applies Flow Field 1 to the moving volume, followed by Flow Field 2. In contrast, the bottom branch (\textit{yellow}) illustrates the DSTN pipeline, where the two flow fields are fused before being applied to the moving volume.}
    \label{fig:DSTN pipeline}
\end{figure}

However, this method requires two rounds of resampling with interpolation, potentially blends neighboring values and blurs edge-rich structures such as dense tissue. To mitigate this problem, we first approximate the two-round transformation and interpolation by composing the two transformations, and then perform a single interpolation named as DSTN. This approximation can be formally expressed as:

\begin{equation}\label{Two-stage STN equation 2}
 m \circ \phi \approx \sum_{\mathbf{q} \in \mathbb{Z}^3(\mathbf{p}_{final})} m(\mathbf{q}) \prod_{d \in \{x,y,z\}} (1 - |\mathbf{p}_{final, d} - \mathbf{q}_d|)
\end{equation}
where $\mathbf{p}_{final} = \mathbf{p} + \mathbf{u}_{volume}(\mathbf{p}) + \mathbf{u}_{mask}(\mathbf{p} + \mathbf{u}_{volume}(\mathbf{p}))$, representing applying both $\mathbf{u}_{volume}$ and $\mathbf{u}_{mask}$ to the original coordinate $\mathbf{p}$. Fig. \ref{fig:DSTN pipeline} demonstrate the difference between DSTN and directly applying STN twice. The effectiveness of DSTN has been empirically and visually validated through a comparison between training with the two-round STN and with DSTN (see Sec. \ref{sec:Ablation Study}).

\subsection{EDT-Based Warping Algorithm}\label{sec:methods:distance transformation}
During the implementation of our registration framework, we observed that conventional warping techniques—such as nearest-neighbor interpolation—tend to degrade fine structural details in binary masks, as demonstrated in Fig. \ref{fig:EDT_result}. This issue was notably absent when the same interpolation methods were applied to breast MRI images. We believe that such interpolation schemes inherently smooth binary structures, leading to the loss of high-frequency boundary information—a concern previously highlighted in other literature \cite{jayroe1976nearest,wang2010improved}. To mitigate this, our approach transforms the binary mask into a signed distance map using a EDT \cite{frisken2000adaptively} prior to warping. This produces a continuous-valued representation that preserves structural fidelity throughout the deformation process. 

The signed distance at each voxel is defined as the difference between its distance to the nearest background and foreground boundary. Let \( M: \Omega \subset \mathbb{R}^3 \rightarrow \{0, 1\} \) denote a binary mask defined on a volumetric domain \( \Omega \), where \( M(x) = 1 \) indicates foreground (object) voxels and \( M(x) = 0 \) indicates background. The signed distance transform \( D: \Omega \rightarrow \mathbb{R} \) assigns to each voxel \( x \in \Omega \) a value equal to its Euclidean distance to the nearest boundary voxel, with a sign that indicates whether \( x \) lies inside or outside the object. Formally, it is defined as \eqref{distance_transform}
\begin{equation}\label{distance_transform}
D(x) = \text{dist}_{\text{edt}}(x, \partial M^+) - \text{dist}_{\text{edt}}(x, \partial M^-),
\end{equation}
where \( \partial M^+ \) and \( \partial M^- \) denote the boundaries of the foreground and background regions, respectively. 

The function \( \text{dist}_{\text{edt}}(x, S) \) denotes the Euclidean distance from voxel \( x \) to the nearest point in the set \( S \subset \Omega \), defined in eq. \eqref{edt}
\begin{equation}\label{edt}
    \text{dist}_{\text{edt}}(x, S) = \min_{y \in S} \|x - y\|_2
\end{equation}

After converting the binary mask into a signed distance map, we apply the deformation field using a spatial transformer with bilinear interpolation, followed by re-binarization of the warped distance map using zero as the threshold to recover the final mask. The pipeline for this algorithm is shown at the bottom left of Fig. \ref{fig:pipleline}.The effectiveness of this algorithm is demonstrated both qualitatively and quantitatively in Sec. \ref{sec:ablation:EDT}.

\subsection{Loss Function}
\label{sec:methods:loss}
The loss function for GuidedMorph consists of three components: volume similarity, local region similarity, and regularization for penalizing fused spatial variations in $\phi$. Prior registration loss functions typically penalize differences between fixed masks and warped moving masks \cite{xu2019deepatlas,li2024semi,voxelmorph}, yet we found that these differences often exhibit discontinuities and fluctuate sharply with even small transformations. Instead, we penalize the differences between the volumes under the masks, offering a smoother and more stable optimization process. Our new loss function is then defined as

\begin{equation}\label{loss_general}
\begin{aligned}
\mathcal{L}(f, f_{loc}, m, m_{loc}, \phi) &= L_{\text{sim}}(f, m \circ \phi) + \lambda L_{\text{smooth}}(\phi) \\
&\quad + \alpha L_{\text{sim}}(f_{loc}, m_{loc} \circ \phi) 
\end{aligned}
\end{equation}
Where the masked fixed and moving volumes are $f_{loc}$ and $m_{loc}$, respectively. $\lambda$ and $\alpha$ are smoothness regularization and volume similarity strength hyperparameters, respectively. The two most common similarity functions for $L_{\text{sim}}$ are the mean square difference (Eq. \eqref{mse}) and local cross correlation (Eq. \eqref{cc}), defined as
\begin{equation}\label{mse}
    MSE(f,m\circ \phi) = \frac{1}{|\Omega|}\sum_{p\in\Omega}[f(p) - [m\circ \phi](p)]^2
\end{equation}
and

\begin{equation}\label{cc}
CC(f, m \circ \phi) =
\frac{\sum_{\mathbf{p} \in \Omega} \left( N(\mathbf{p}) \right)^2}
{\sum_{\mathbf{p} \in \Omega} \left( \sum_{\mathbf{p}_i} \bar{f}(\mathbf{p}_i)^2 \right)
\left( \sum_{\mathbf{p}_i} \bar{m}(\mathbf{p}_i)^2 \right)}
\end{equation}
respectively, where $\bar{f}(\mathbf{p}_i)$, $\bar{m}(\mathbf{p}_i)$, and $N(\mathbf{p})$ are defined in \eqref{cc_a}, \eqref{cc_b}, and \eqref{cc_c}. 
\begin{equation}\label{cc_a}
    \bar{f}(\mathbf{p}_i) = f(\mathbf{p}_i) - \hat{f}(\mathbf{p})
\end{equation}
\begin{equation}\label{cc_b}
    \bar{m}(\mathbf{p}_i) = (m \circ \phi)(\mathbf{p}_i) - (\hat{m} \circ \phi)(\mathbf{p})
\end{equation}
\begin{equation}\label{cc_c}
    N(\mathbf{p}) = \sum_{\mathbf{p}_i} \bar{f}(\mathbf{p}_i) \, \bar{m}(\mathbf{p}_i)
\end{equation}

\section{Dataset}
\subsection{ISPY2 Dataset}
The ISPY2 dataset is a public dataset introduced by \cite{wang2019spy} that contains Dynamic Contrast-Enhanced (DCE) breast MRI data for 711 patients diagnosed with early-stage breast cancer. These patients were adaptively randomized to drug treatment arms between 2010 and 2016. The images, acquired from over 22 clinical centers, focus on patients undergoing neoadjuvant chemotherapy (NAC). Each patient underwent four MRI exams at specific timepoints: pretreatment (``T0''), early-treatment (``T1''), mid-treatment (``T2''), and post-treatment (``T3'') during their NAC.

For this study, we randomly selected 100 patients from the ISPY2 dataset, each with four pre-contrast breast MRI studies. This subset was divided into training and test sets,consisting of 80 and 20 patients, respectively. To simulate a clinical scenario in which physicians monitor sequential changes across multiple examinations, we constructed volume pairs for registration using the timepoints (``T0'', ``T1''), (``T1'', ``T2''), and (``T2'', ``T3''). This approach resulted in a total of 240, and 60 pairs for the training, and test sets, respectively.

Before use, all volumes were resampled to a $128\times 256\times 256$ grid with 1 mm isotropic voxels. Data preprocessing included (1) clipping intensities to the 1st and 99th percentiles to minimize outlier effects, and (2) linearly scaling these values to a 0–1 range. Additionally, affine volume registration was performed using the Advanced Normalization Tools (ANTS).

\subsection{Internal Breast Dataset}
For the internal dataset, we collected breast MRI scans from our institution (The research protocol was approved by the Health System Institutional Review Board (IRB) Pro00112340), spanning from March 2014 to December 2021. To align with the dataset selected from ISPY2, we randomly selected 100 patients, each with four pre-contrast breast MRI studies. This subset was divided into training and test sets, consisting of 80 and 20 patients, respectively. Furthermore, the four studies were ordered by the study date to simulate the ``T0,'' ``T1,'' ``T2,'' and ``T3'' timepoints in the ISPY2 dataset, and registration pairs were constructed accordingly. This approach resulted in a total of 240 and 60 pairs for the training and test sets, respectively. The preprocessing steps were identical to those applied to the ISPY2 dataset.

\subsection{Mask Generation}\label{segmentation}
To generate masks for the ISPY2 and internal dataset, we followed \cite{lew2024publicly,chen2025breast} to train a patch-based VNet \cite{milletari2016v} on the Duke Breast MRI dataset \cite{saha2018machine,lew2024publicly}. The patch size is set to $96$ and we utilize sliding window-based inference. We augment the data with random brightness changes. The model achieves 0.95 and 0.88 Dice segmentation coefficient (DSC) on the breast and dense tissue respectively, and was selected by validation performance.

\section{Experiments}
\subsection{Implementation}\label{Implementation section}
The proposed deformable registration method was developed using the PyTorch library in Python. Our method is built on a 3D convolutional neural network (CNN) which we detail later in this section, trained at an initial learning rate of $5 \times 10^{-4}$. To achieve stable optimization, we employed the AdamW optimizer with momentum, using a batch size of 1. 

To adjust the balance between similarity penalty and regularization across different loss functions (Eq. \ref{loss_general}), we chose the hyper-parameter settings as follows. For the MSE loss function, which has a higher similarity penalty, we set $\alpha = 1$ and $\lambda = 0.08$ to ensure sufficient regularization. Conversely, for the CC loss function where the similarity penalty is lower, we adjusted the weights to $\alpha = 1$ and $\lambda = 8$, thus maintaining a consistent balance between the similarity penalty and regularization across both metrics. The model was trained on a single 48GB NVIDIA RTX A6000 GPU for 500 epochs, the CPU running times are evaluated on AMD Ryzen Threadripper 3970X, with 32 cores, 64 threads, and 128 MB L3 cache. Detailed information on the 3D CNN architecture, and dynamic loss functions can be found in the subsequent subsections. Since training with loss functions based on MSE and CC yields comparable performance, as shown in Sec. \ref{sec:Ablation Study}, in the experiment, we conduct the baseline implementation and result comparison using only MSE for faster convergence.

\subsubsection{3D CNN Architecture}
We adopt the CNN architecture from both VoxelMorph \cite{voxelmorph} and TransMorph \cite{chen2022transmorph} for our GuidedMorph (Voxel) and GuidedMorph (Trans), respectively. For GuidedMorph (Voxel), we follow the implementation in \cite{voxelmorph}, using a kernel size of 3 in both the encoder and decoder stages, and a stride of 2. Each convolution is followed by a LeakyReLU activation with a slope of 0.2 to introduce non-linearity. In the decoding phase, the network alternates between upsampling and deconvolution layers, incorporating skip connections that directly pass features from the encoder to the corresponding layers involved in generating the registration. The innermost layer performs convolutions on a volume of size $(1/16)^3$ relative to the input dimensions.

For GuidedMorph (Trans), we adopt the Swin Transformer-based architecture introduced in \cite{chen2022transmorph}. The network begins by dividing the input volumes into non-overlapping 3D patches of size $2 \times P \times P$, with $P$ typically set to 4, resulting in a patch embedding of $8C$ dimensions. These patches are flattened and linearly projected into token embeddings of dimension $C = 96$. The tokens pass through four sequential stages of Swin Transformer blocks, each operating at decreasing spatial resolutions and increasing feature depths. To recover spatial precision lost during tokenization, we include two convolutional paths—one using the original resolution and the other using a downsampled version of the input. Each path extracts high-resolution local features, which are concatenated with the Transformer outputs and passed to the decoder through skip connections. The decoder comprises successive upsampling layers with $3 \times 3 \times 3$ convolutions and LeakyReLU activations (slope 0.2). The output deformation field is generated by a final convolutional layer and used by the spatial transformer module to warp the moving image.

\subsubsection{Dense Tissue Extraction Strategy}\label{sec:experiments:feature_extraction}
\begin{figure}[htbp]
    \centering
    \includegraphics[width=1\linewidth]{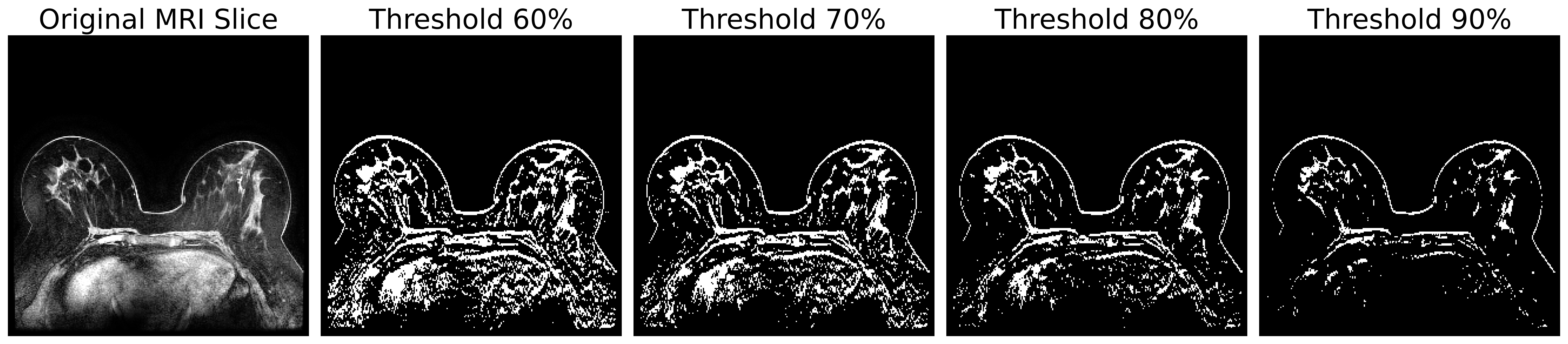}
    \caption{Visualization of the threshold step across different percentile thresholds.}
    \label{fig:threshold}
\end{figure}

As mentioned in Sec. \ref{sec:methods:selection strategy}, this section is utilized to demonstrate the technical implementation details of our filter-based mask extraction algorithm, specifically the threshold selection to reduce noise and better resemble the true dense tissue mask. As shown in Fig. \ref{fig:threshold}, the noise is quite obvious even when we set the threshold up to 60 percentile. With the increase of the threshold, there is an obvious noise reduction, but it also lead to the loss of certain detailed information in dense tissue. To further confirm which threshold provides the best performance on our task, we conduct experiments on our GuidedMorph (Trans), with thresholds for filter-based dense tissue extraction ranging from 60 to 90 percentile. The performance is evaluated using the Dice coefficient between the registered source and target dense tissue masks. Since the result shows that a threshold of 90 yields the best performance, so we adopt it in our filter-based dense tissue extraction.

\subsubsection{Dynamic Loss Functions} 
To achieve the dynamic region focus we mentioned in the Sec. \ref{sec:methods:overview}, we double the $\alpha$ parameter every 100 epochs for both MSE and CC similarity settings to enhance the model's focus on aligning masked critical sub-regions. This approach not only ensures alignment across the entire structure during the early stages of training but also guarantees that the network thoroughly captures and enhances details within the dense tissue as training progresses.

\subsection{Baseline methods}
For comparison, we selected two traditional optimization-based registration algorithms, \textbf{SyN} and \textbf{Nifty-reg} \cite{avants2008symmetric,modat2010fast} and three learning-based registration algorithms including \textbf{VoxelMorph}, \textbf{CycleMorph} and \textbf{TransMorph} \cite{voxelmorph,kim2021cyclemorph,chen2022transmorph} due to their superior performance in previous registration tasks \cite{voxelmorph, chen2022transmorph}. These were chosen as representatives for previous registration algorithms that have demonstrated state-of-the-art registration performance. While VoxelMorph and TransMorph propose multiple architectures, including $\textit{-diff}$, $\textit{-Bayes}$, and $\textit{-bspl}$, we chose the architectures that have demonstrated the best performance on brain label Dice coefficients in their respective works \cite{voxelmorph, chen2022transmorph}—namely, "VoxelMorph-2" (denoted as "VolumeMorph" hereafter) and "TransMorph."

For the hyper-parameters, we adhered to the default hyper-parameter settings as specified in their respective publications or GitHub repositories. To ensure a fair comparison, all preprocessing methods applied to our GuidedMorph model were also uniformly implemented across these baseline methods. To demonstrate the effectiveness of two-stage registration and eliminate the effect given by the extra segmentation mask input, we also built single-stage network incorporates the segmentation masks. The network is denoted as ``TransMorph + DT'' for those using a dense tissue mask and ``TransMorph + B'' for those using breast mask in the result table VI. The TransMorph is utilize in the experiment since it demonstrates the best performance among all deep-learning-based baselines. 

The only difference between ``TransMorph + DT'' and ``Ours + DT'' is that ``Ours + DT'''s flow filed is the composition generated by two stages of the network, whereas ``TransMorph + DT'' uses the flow field solely from the first stage.  The same distinction applies to ``TransMorph + B'' and ``Ours + B''. All other settings, including hyperparameters, loss functions, and input variables, are kept the same to ensure a fair comparison. 
\subsection{Evaluation}\label{Evaluation section}
Due to the high accuracy of the segmentation algorithm-validated to achieve a Dice coefficient exceeding 88\% for breast dense tissue and 95\% for overall breast- we are able to directly obtain dense tissue and breast masks for the test dataset. Using these masks, we evaluate the registration performance by comparing the deformed segmentation map with the fixed mask, employing the Dice coefficient score. The method for calculating this score is detailed in Eq. \eqref{dice}.
\begin{equation}\label{dice}
Dice(f_{mask}, m_{mask} \circ \phi) = 2 \cdot \frac{|f_{mask} \cap (m_{mask} \circ \phi)|}{|f_{mask}| + |m_{mask} \circ \phi|}
\end{equation}
The Dice coefficient approaching 1 indicates that the registered moving volume and the fixed volume are perfectly aligned, while a Dice coefficient of 0 demonstrates no overlap. To further eliminate potential bias from the evaluation process given by segmentation model, we also include the SSIM measurements on the breast (excluding the chest part due to our task focus) as an alternative similarity metric for the registered moving volume and the fixed volume shown in Eq. \eqref{SSIM}.
\begin{equation}\label{SSIM}
    SSIM(f, m \circ \phi) = \frac{(2\mu_x \mu_y + c_1)(2\sigma_{xy} + c_2)}{(\mu_x^2 + \mu_y^2 + c_1)(\sigma_x^2 + \sigma_y^2 + c_2)}
\end{equation}
Where the $\mu_x, \mu_y$ are the average intensities of volume $f$ and $ m \circ \phi$, $\sigma^2_x, \sigma^2_y$ are the variances and $\sigma_{xy}$ is the covariance of two volume $f$ and $m \circ \phi$. $c_1$ and $c_2$ are the constants used to stabilize the division with weak denominator.

To measure the consistency of the deformation fields, we presented the percentages of non-positive values in the determinant of the Jacobian matrix associated with the deformation fields. In the result Sec. \ref{result} the matrix is denoted as \% of $|J_{\phi}|\leq 0$. Additionally, we have displayed the running times for each method to demonstrate their efficiency.

\section{Main results}\label{result}
The results section includes both qualitative and quantitative evaluation result for the breast registration on ISPY2 and our internal breast dataset. While loss function based on CC and MSE gives comparable performance shown in \ref{sec:ablation:loss}, the result shown in the below sections are evaluated with MSE. Since TransMorph demonstrates the best performance among all learning-based baselines, we primarily evaluate our GuidedMorph (Trans) in both qualitative and quantitative analyses. Additionally, we include a comparison between the single-stage baseline and our GuidedMorph (Voxel) in Tab. \ref{tab:result_table_voxel} on the ISPY2 dataset.
\subsection{Qualitative Evaluation}
\begin{figure*}[htbp]
    \centering
    \includegraphics[width=1\linewidth]{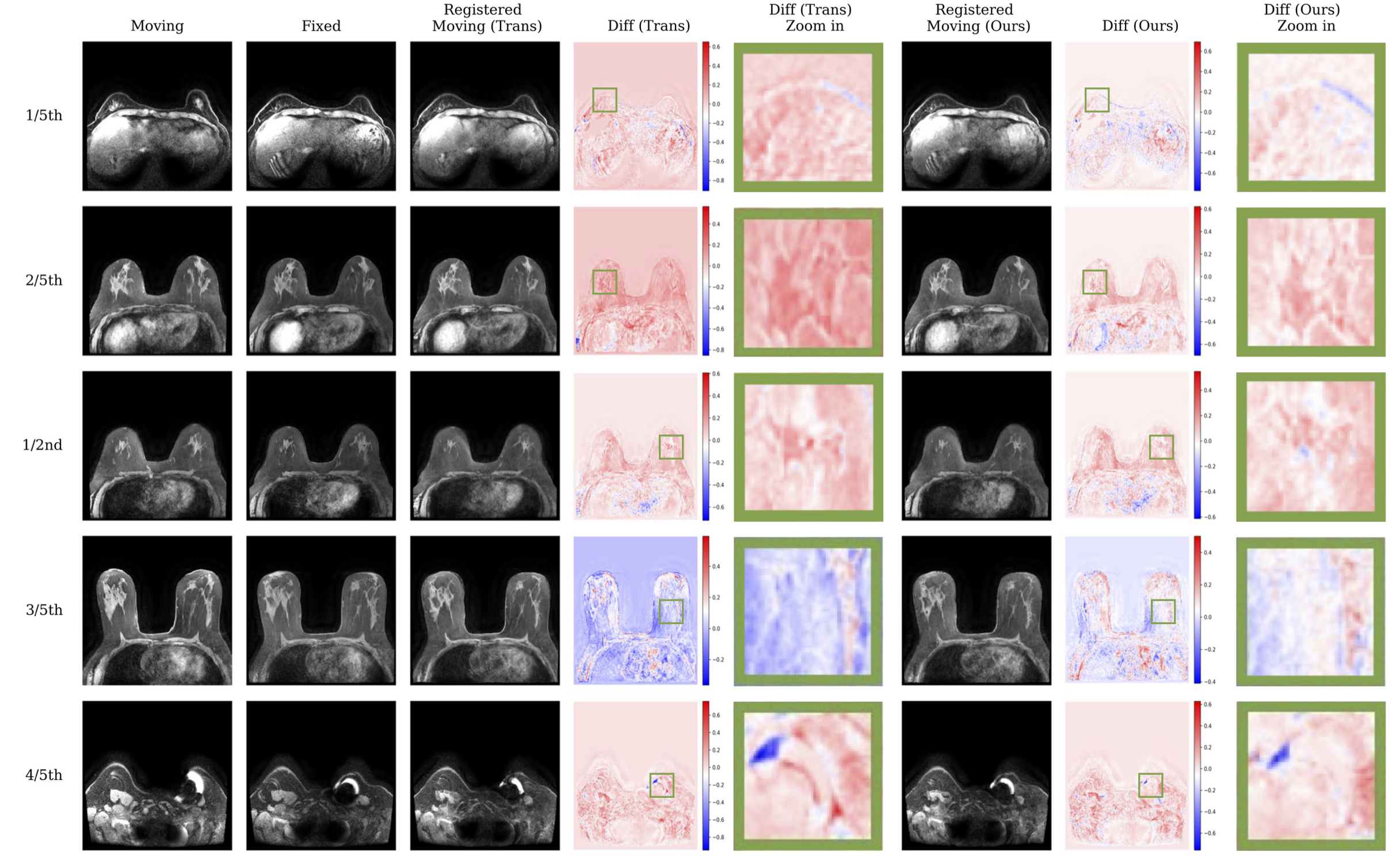}
    \caption{\textbf{Qualitative comparison of our model versus the TransMorph baseline:} Rows, from top to bottom, show slices at the 1/5th, 2/5th, 1/2nd, 3/5th, and 4/5th positions from different MRI volumes. Columns, from left to right, display the source slice, target slice, Trans-registered slice, Trans-target difference, green box zoom-in from Trans-target difference, our registered slice, our model’s difference, green box zoom-in from our model’s difference. The green boxes highlight regions where our method achieves better alignment, and the zoom-in columns offer enhanced visualization of these areas.}
    \label{fig:result_figure}
\end{figure*}

Fig. \ref{fig:result_figure} presents the qualitative evaluation of GuidedMorph (Ours), based on TransMorph architecture, and filter-based dense tissue extraction method, compared against TransMorph on ISPY2 and our internal breast dataset, respectively. For each result figure, rows from top to bottom show slices at the 1/5th, 2/5th, 1/2nd, 3/5th, and 4/5th positions from different MRI volumes. Columns, from left to right, display the source
slice, target slice, Trans-registered slice, Trans-target difference, green box zoom-in from Trans-target difference, our registered
slice, our model’s difference, green box zoom-in from our model’s difference. While our model’s differences show lighter colors (indicating smaller discrepancies) compared to the Trans-target difference, the results demonstrate the overall superiority of our model. We also add green boxes in the result figure to highlight regions within the images, emphasizing the improved alignment of dense tissue achieved by GuidedMorph (Trans) over TransMorph. Compared to the general difference color, the green boxes emphasize more on the dense tissue area due to the focus of our study.

\subsection{Quantitative Evaluation}
To provide a comprehensive comparison of GuidedMorph with other established registration methods, Table \ref{tab:result_table} details the quantitative evaluation metrics used on ISPY2 and the internal breast dataset, including the Dice score for dense tissue (Dice$_{DT}$), Dice score for breast tissue (Dice$_{B}$), SSIM for breast, the percentage of non-positive values in the Jacobian determinant, and the runtime on both GPU and CPU platforms. Detailed descriptions of the evaluation metrics and algorithm implementations are presented in Sec. \ref{Evaluation section} and \ref{Implementation section}, respectively. 

The results demonstrate that GuidedMorph, in its filter-based mode without any segmentation masks input (Our (Filter)), achieves higher Dice$_{DT}$, Dice$_{B}$, and SSMI compared to other registration algorithms that do not rely on segmentation models. The percenetage of non-positive values in the Jacobian determinant for the Our model is comparable to that of previous state-of-the-art methods. When utilizing segmentation masks for both dense and breast tissues (Ours + DT and Ours + B), our GuidedMorph further surpasses the performance of single-layer TransMorph equipped with the same masks in terms of Dice scores, SSIM, and the stability of the transformation as indicated by the lower percenetage of non-positive values in the Jacobian determinant. 

Specifically, averaged across both datasets, our (Filter) achieves an improvement of \textbf{12.2\%} in Dice$_{DT}$, \textbf{5.51\%} in Dice$_{B}$, and \textbf{1.38\%} in SSIM compared to the best learning–based baseline that do not rely on segmentation models. Compared to the best performing optimization-based algorithm, our model significantly reduces the inference time and improves Dice$_{DT}$ by \textbf{5.29\%} while maintaining comparable Dice$_{B}$ and breast SSIM.

For our model with segmentation mask input, we observe a \textbf{21.16\%} improvement in dense tissue Dice coefficient, \textbf{3.13\%} in breast Dice coefficient, and \textbf{2.11\%} in overall breast SSIM compared to the single stage baseline with same mask input.

We also demonstrate that our framework is compatible not only with TransMorph but also with VoxelMorph as the single-stage backbone, by comparing filter-based GuidedMorph (Voxel) with the original VoxelMorph. As a result, our model achieves improvements of \textbf{13.01\%} in Dice${DT}$, \textbf{4.91\%} in Dice${B}$, and \textbf{1.21\%} in SSIM over VoxelMorph.

\begin{table*}[h]
\label{tab:result_table}
\caption{Quantitative comparison between our GuidedMorph (Trans), and baseline registration models on the ISPY2 and internal dataset under three evaluation settings} 
\centering
\resizebox{\linewidth}{!}{
\begin{tabular}{l|c|c|c|c|c|c|c}
\hline
\multicolumn{8}{c}{\textbf{ISPY2 Dataset}} \\
\hline
Method & \textbf{Dice$_{DT}$} & Dice$_{B}$ & SSIM & \% of $|J_{\phi}| \leq 0$ & GPU sec & CPU sec & Seg.\\ 
\hline\hline
Affine & 0.467 $\pm$ 0.136 & 0.850 $\pm$ 0.062 & 0.835 $\pm$ 0.053 & - & - & - & $\times$\\
SyN & 0.686 $\pm$ 0.135 & 0.907 $\pm$ 0.049 & 0.885 $\pm$ 0.039 & 0.006 $\pm$ 0.018 & - & 255.411 $\pm$ 34.201 & $\times$\\
NiftyReg & 0.718 $\pm$ 0.128 & 0.908 $\pm$ 0.050 & 0.897 $\pm$ 0.037 & 0.213 $\pm$ 0.211 & - & 45.011 $\pm$ 9.594 & $\times$ \\
VoxelMorph & 0.672 $\pm$ 0.127 & 0.881 $\pm$ 0.059 & 0.891 $\pm$ 0.035 & 2.244 $\pm$ 0.962 &  0.106 $\pm$ 0.025 & 3.577 $\pm$ 0.410 & $\times$ \\
CycleMorph & 0.670 $\pm$ 0.126 & 0.881 $\pm$ 0.059 & 0.886 $\pm$ 0.036 & 2.387 $\pm$ 0.921 & 0.095 $\pm$ 0.026 & 3.224 $\pm$ 0.252 & $\times$ \\
TransMorph & 0.681 $\pm$ 0.126 & 0.883 $\pm$ 0.058 & 0.894 $\pm$ 0.034 & 2.138 $\pm$ 0.981 & 0.208 $\pm$ 0.027 & 5.799 $\pm$ 1.255 & $\times$ \\
TransMorph + B & 0.706 $\pm$ 0.124 &  0.912 $\pm$ 0.050 & 0.902 $\pm$ 0.033 & 0.332 $\pm$ 0.248 & 0.210 $\pm$ 0.041 & 5.818 $\pm$ 0.157 & \checkmark \\
TransMorph + DT & 0.725 $\pm$ 0.118 & 0.904 $\pm$ 0.054 & 0.897 $\pm$ 0.035 & 0.434 $\pm$ 0.361 & 0.211 $\pm$ 0.047 & 5.854 $\pm$ 0.138 & \checkmark \\\hline
Ours (Filter) & 0.756 $\pm$ 0.120 & 0.908 $\pm$ 0.050 & 0.904 $\pm$ 0.033 & 0.783 $\pm$ 0.430 &  0.412 $\pm$ 0.046 & 11.003 $\pm$ 0.147 & $\times$ \\
Ours + B & 0.738 $\pm$ 0.119 & \textbf{0.945 $\pm$ 0.041 }& \textbf{0.938 $\pm$ 0.024} & 0.338 $\pm$ 0.288 & 0.416 $\pm$ 0.043 & 11.203 $\pm$ 0.175 & \checkmark \\
Ours + DT & \textbf{0.934 $\pm$ 0.062} & 0.908 $\pm$ 0.052 & 0.899 $\pm$ 0.034 & 0.366 $\pm$ 0.281 & 0.420 $\pm$ 0.047 & 11.192 $\pm$ 0.157 & \checkmark \\\hline
\multicolumn{7}{c}{\textbf{Internal Dataset}} \\
\hline
Method & \textbf{Dice}$_{DT}$ & Dice$_{B}$ & SSIM & $\% \text{ of } |J_{\phi}| \leq 0$ & GPU sec & CPU sec & Seg.\\
\hline\hline
Affine & 0.402 $\pm$ 0.225 & 0.772 $\pm$ 0.123 & 0.878 $\pm$.0561 & - & - & - & $\times$ \\
SyN & 0.578 $\pm$ 0.233 & 0.855 $\pm$ 0.099 & 0.905 $\pm$ 0.051 & 0.005 $\pm$ 0.008 & - & 257.039 $\pm$ 69.360 & $\times$ \\
NiftyReg & 0.612 $\pm$ 0.223 & 0.876 $\pm$ 0.071 & 0.915 $\pm$ 0.046 & 0.339 $\pm$ 0.442 & - & 36.106 $\pm$ 8.792 & $\times$ \\
VoxelMorph & 0.584 $\pm$ 0.209 & 0.814 $\pm$ 0.112 & 0.911 $\pm$ 0.042 & 3.225 $\pm$ 1.859 & 0.103 $\pm$ 0.025 & 3.492 $\pm$ 0.576 & $\times$ \\
CycleMorph & 0.581 $\pm$ 0.211 & 0.819 $\pm$ 0.111 & 0.908 $\pm$ 0.044 & 3.600 $\pm$ 1.658 & 0.094 $\pm$ 0.057 &  2.505 $\pm$ 0.273 & $\times$ \\
TransMorph & 0.600 $\pm$ 0.204 & 0.818 $\pm$ 0.110 & 0.909 $\pm$ 0.042 & 3.438 $\pm$ 1.873 & 0.212 $\pm$ 0.053 & 4.945 $\pm$ 0.087 & $\times$ \\
TransMorph + B & 0.609 $\pm$ 0.214 & 0.888 $\pm$ 0.077 & 0.919 $\pm$ 0.040 & 0.864 $\pm$ 0.638 & 0.222 $\pm$ 0.059 & 6.326 $\pm$ 0.589 & \checkmark \\
TransMorph + DT & 0.640 $\pm$ 0.207 & 0.849 $\pm$ 0.104 & 0.915 $\pm$ 0.042 & 1.027 $\pm$ 0.704 & 0.222 $\pm$ 0.056 & 6.487 $\pm$ 0.532 & \checkmark \\\hline
Ours (Filter) & 0.680 $\pm$ 0.201 & 0.885 $\pm$ 0.103 & 0.924 $\pm$ 0.038 & 1.472 $\pm$ 1.126 &  0.412 $\pm$ 0.047 & 10.022 $\pm$ 0.184 & $\times$ \\
Ours + B & 0.690 $\pm$ 0.192 & \textbf{0.942 $\pm$ 0.055} & \textbf{0.954 $\pm$ 0.027} & 0.656 $\pm$ 0.539 & 0.412 $\pm$ 0.054 & 10.373 $\pm$ 0.194 & \checkmark \\
Ours + DT & \textbf{0.883 $\pm$ 0.136} & 0.869 $\pm$ 0.092 & 0.919 $\pm$ 0.042 & 0.751 $\pm$ 0.613 & 0.413 $\pm$ 0.058 & 10.546 $\pm$ 0.191 & \checkmark \\\hline
\multicolumn{8}{p{\linewidth}}{(1) without additional segmentation model (Filter), (2) with a breast segmentation mask as auxiliary input (+B), and (3) with a dense tissue segmentation mask as auxiliary input (+DT). For each setting, we report the Dice coefficient for dense tissue (Dice$_{DT}$, our primary evaluation metric), overall breast Dice (Dice$_{B}$), structural similarity index (SSIM), percentage of non-positive Jacobian determinant (\% of $|J_{\phi}| \leq 0$), and runtime (GPU and CPU). The best performance in each setting is highlighted in bold, and Dice$_{DT}$ is bolded to emphasize its importance in this study. “Seg.” indicates whether the method requires an external segmentation model as input (\checkmark) or not ($\times$).}\\
\end{tabular}}
\label{tab:result_table}
\end{table*}

\begin{table*}[h]
\centering
\caption{Quantitative comparison between our GuidedMorph (Voxel) in the filter-based setting and the baseline VoxelMorph model on the ISPY2 dataset} 
\resizebox{\linewidth}{!}{
\begin{tabular}{l|c|c|c|c|c|c|c}
\hline
\multicolumn{8}{c}{\textbf{ISPY2 Dataset}} \\
\hline
Method & \textbf{Dice}$_{DT}$ & Dice$_{B}$ & SSIM & \% of $|J_{\phi}| \leq 0$ & GPU sec & CPU sec & Seg.\\ 
\hline\hline
VoxelMorph & 0.672 $\pm$ 0.127 & 0.881 $\pm$ 0.059 & 0.891 $\pm$ 0.035 & 2.244 $\pm$ 0.962 &  0.106 $\pm$ 0.025 & 3.577 $\pm$ 0.410 & $\times$ \\
Ours (Filter) & 0.660 $\pm$ 0.204 & 0.854 $\pm$ 0.098 & 0.922 $\pm$ 0.038 & 1.284 $\pm$ 0.921 & 0.215 $\pm$ 0.066 & 6.660 $\pm$ 0.195 & $\times$ \\\hline
\end{tabular}}
\label{tab:result_table_voxel}
\end{table*}

\section{Ablation Studies} \label{sec:Ablation Study}
\begin{table}[t]
\label{tab:ablation_table1}
\caption{Hyperparameter $\alpha$ analysis}
\centering
\begin{tabular}{c|c}
$\alpha$ & Dice$_{DT}$ \\
\hline
$0.01$ & 0.755 $\pm$ 0.120 \\
$0.1$ & 0.757 $\pm$ 0.122 \\
$1$ & 0.756 $\pm$ 0.120 \\
$10$ & 0.746 $\pm$ 0.121 \\
$100$ & 0.732 $\pm$ 0.128 \\
\end{tabular}
\label{tab:ablation_table1}
\end{table}

\begin{table}[t]
\caption{Variation of module}
\centering
\begin{tabular}{c|c}
Ablation & Dice$_{DT}$ \\
\hline
First Stage Only/ TransMorph (Filter) & 0.705 $\pm$ 0.126 \\
Second Stage Only & 0.719 $\pm$ 0.124 \\
STN Twice & 0.735 $\pm$ 0.123 \\
Our (Filter) & 0.756 $\pm$ 0.120 \\
\end{tabular}
\label{tab:ablation_table2}
\end{table}

\begin{table}[t]
\caption{Mask warping algorithm}
\centering
\begin{tabular}{c|c}
Method (EDT + Interp) & Dice$_{DT}$ \\
\hline
no EDT (Bilinear) & 0.455 $\pm$ 0.143 \\
no EDT (Nearest-Neighbor) & 0.738 $\pm$ 0.119 \\
with EDT (Nearest-Neighbor) & 0.740 $\pm$ 0.119 \\
with EDT (Bilinear) & 0.756 $\pm$ 0.120 \\
\end{tabular}
\label{tab:ablation_table4}
\end{table}

\begin{table}[t]
\caption{Loss function on performance}
\centering
\begin{tabular}{c|c}
Loss Type & Dice$_{DT}$ \\
\hline
MAE & 0.756 $\pm$ 0.120 \\
CC & 0.757 $\pm$ 0.118 \\
\end{tabular}
\label{tab:ablation_table3}
\end{table}

In this section, we design experiments to evaluate the components and choices made in our proposed registration model in Sec. \ref{sec:Method}. All experiments are conducted using our GuidedMorph (Trans) in the filter-based setting on the ISPY2 dataset. The study is divided into five key parts:

(1) \textbf{Hyper-parameter Analysis} demonstrates the performance of TransMorph and filter-based GuidedMorph (Trans) under a varied range of hyper-parameters.

(2) \textbf{Network Components} presents two additional experiments to demonstrate the efficacy of the first and second stages in our registration model, as described in the GuidedMorph Overview (Section \ref{sec:methods:overview}).

(3) In section \textbf{DSTN Model}, we compare our proposed DSTN with an alternative option that performs the STN twice, as mentioned in Section \ref{sec:methods:twostageSTN}, to demonstrate the effectiveness of DSTN approximation.

(4) \textbf{EDT-based Warping} compares our warping algorithm described in Sec. \ref{sec:methods:distance transformation} with the vanilla warping algorithm using either bilinear or nearest interpolation under various experimental settings.

(5) Finally, in section \textbf{Loss Function}, we investigate the impact of different loss functions on model performance by presenting results using both MSE and NCC. Our findings, summarized in a comparative table, indicate that the choice of loss function does not significantly affect the final outcomes. 

All experiments are conducted on the ISPY2 dataset.
\subsubsection{Hyper-parameter Analysis}
\begin{figure}[htbp] 
    \centering
    \includegraphics[width=1\linewidth]{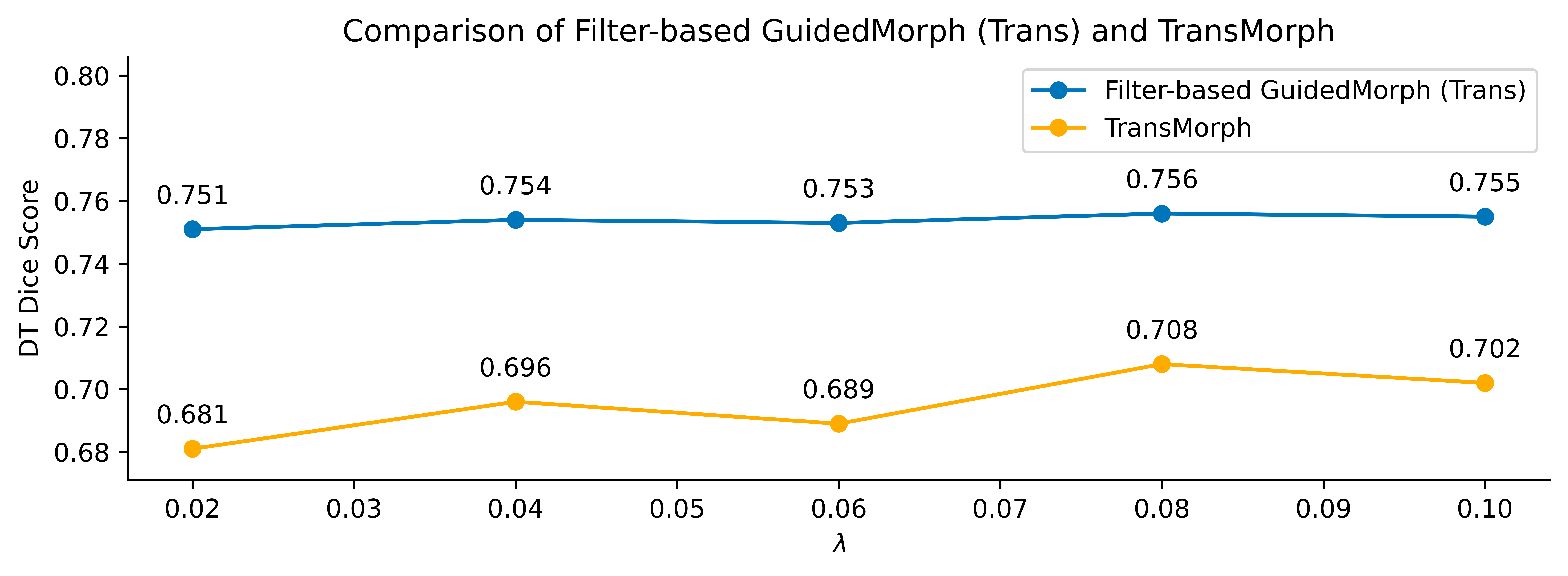}
    \caption{Hyperparameter analysis with comparison between our GuidedMorph ($\textit{blue}$), and baseline TransMorph ($\textit{yellow}$).}
    \label{fig:alpha_ablation} 
\end{figure}

In this section, we evaluate the performance of TransMorph and GuidedMorph (Trans) based on filter-based mask extraction under various values of the hyper-parameters $\alpha$ and $\lambda$, as defined in Eq. \eqref{loss_general}. For $\alpha$, we explore a range of values from 0.02 to 0.10 for both TransMorph  and Our (Filter) based on TransMorph architecture, with the results presented in Figure \ref{fig:alpha_ablation}. Similarly, for $\lambda$, we assess its impact using values of 0.01, 0.1, 1, 10, and 100, the result is shown in Table \ref{tab:ablation_table1}. As a result, our proposed GuidedMorph demonstrates strong stability without significant performance variation across various hyperparameter settings, with the largest p-value of 0.276. Moreover, our model exhibits superior performance compared to TransMorph for all values of $\lambda$, as shown in Fig. \ref{fig:alpha_ablation}. 

\subsubsection{Network Components}
To demonstrate the effectiveness of combining the two stages, in this section we demonstrate the performance of the first stage only and the second stage only, as shown in Tab. \ref{tab:ablation_table2}. For this demonstration, the experiment utilizes the filter-based GuidedMorph (Trans) setting. Notably, the difference between the first-stage-only model and TransMorph is that the former uses the same loss function as GuidedMorph—including crucial-region similarity as one of the polynomial terms. Therefore, the "First Stage Only" model is effectively equivalent to TransMorph (Filter). As a result, combining the two stages improves the performance by \textbf{7.23\%} compared with utilizing the first stage only and by \textbf{5.15\%} compared with utilizing the second stage only.

\subsubsection{DSTN Model}
\begin{figure*}[h] 
    \centering
    \includegraphics[width=1\linewidth]{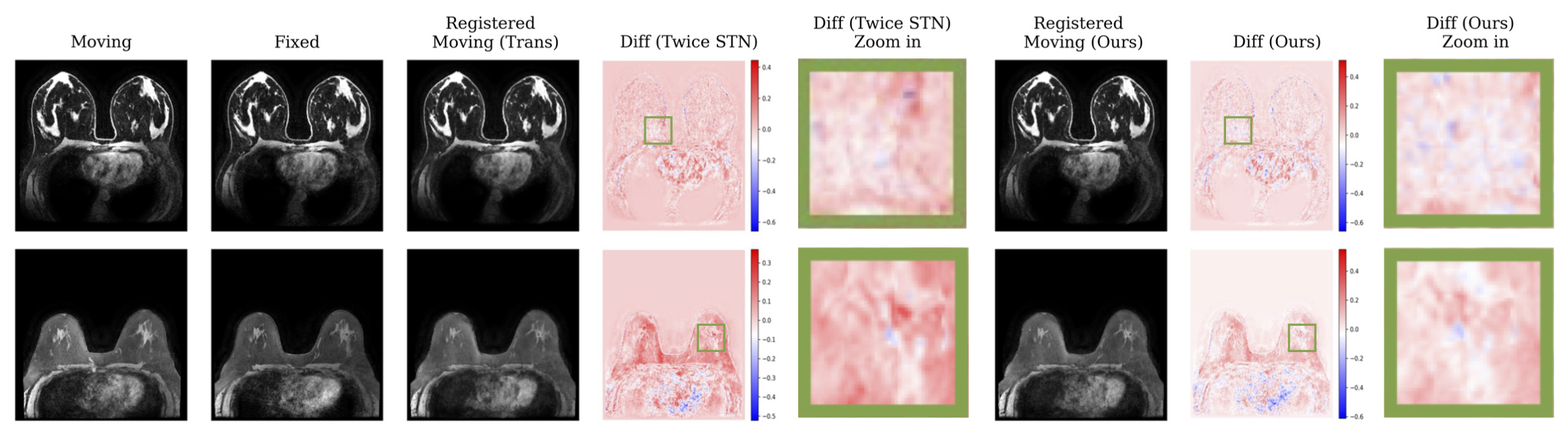}
    \caption{\textbf{Comparison between DSTN and applying STN twice:} Rows show randomly selected slices.}
    \label{fig:DSTN_result} 
\end{figure*}

This section is designed to demonstrate the effectiveness comparing DSTN with applying the STN twice (see Sec. \ref{sec:methods:twostageSTN}). The experiment is also conducted on the filter-based GuidedMorph (Trans) setting, and the final performance evaluated by Dice$_{DT}$ is shown in \ref{tab:ablation_table2}. As shown, the proposed DSTN improved the performance by \textbf{2.86\%}. We also present a visual comparison between the registration results from the two methods, as shown in Fig. \ref{fig:DSTN_result}. Although the differences are not apparent in the registered moving volumes, the difference maps—computed between the fixed and registered moving volumes—generally show lighter colors for DSTN, indicating a smaller gap between our result and the target. The performance differences are especially noticeable in the dense tissue regions, as highlighted by the green boxes.

\subsubsection{EDT-based Warping}\label{sec:ablation:EDT}
\begin{figure*}[h] 
    \centering
    \includegraphics[width=1\linewidth]{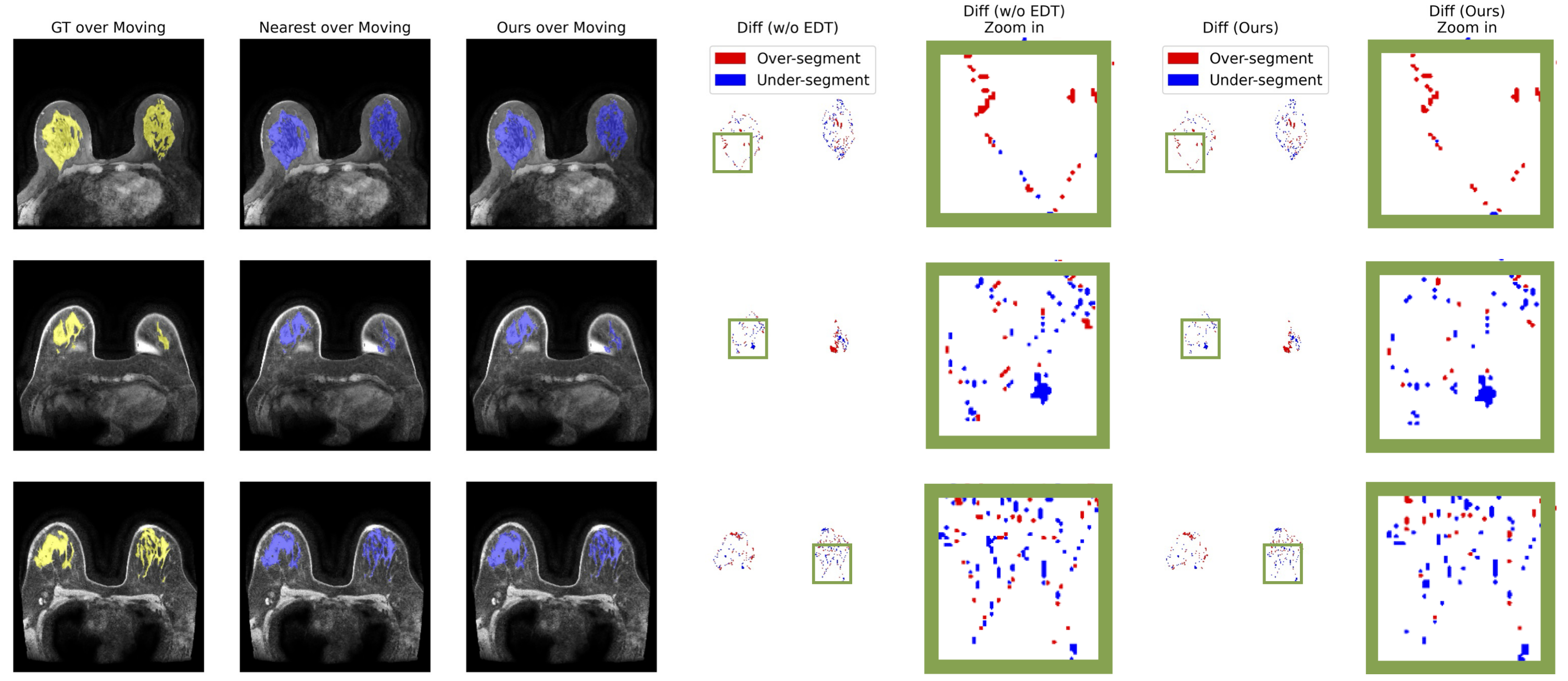}
    \caption{\textbf{Comparison between our method with EDT (ours) and conventional method without EDT (w/o EDT):} Columns, from left to right, show: ground truth (manually modified segmentation, \textit{yellow}) overlaid on the registered moving image, segmentation warped without EDT (nearest interpolation, \textit{blue}), segmentation warped with EDT (\textit{blue}), the difference map between ground truth and no EDT, and its zoomed-in view, followed by the difference map between ground truth and our EDT-based method, and its zoomed-in view. Red and blue indicate over-segmentation and under-segmentation, respectively.}
    \label{fig:EDT_result} 
\end{figure*}

In this section, we aim to evaluate the different mask warping algorithm for our inference mode to prove the effectiveness of our proposed EDT-based Warping method. Tab. \ref{tab:ablation_table4} compares the inference result with and without utilizing the EDT with both bilinear and nearest as interpolation strategy. The result demonstrate that EDT-based Warping method is less sensitive to interpolation algorithm compared to conventional warping algorithm and achieves the improvement of \textbf{2.44\%} compared to warping algorithm without EDT with any interpolation strategy. The visual comparison between our method with EDT (ours) and conventional method without EDT (w/o EDT) is shown in Fig. \ref{fig:EDT_result}. Columns, from left to right, display: the manually modified dense tissue segmentation overlaid on the registered moving slice (ground truth), the nearest-neighbor warped segmentation without EDT, our EDT-warped segmentation, the difference between ground truth and no EDT (with zoomed-in region), and the difference between ground truth and ours (with zoomed-in region). With fewer red and blue points in the difference maps, the results demonstrate that introducing EDT into the mask deformation mitigates both over-segmentation and under-segmentation.

\subsubsection{Loss Function}\label{sec:ablation:loss}
To evaluate the impact of different loss functions on the performance of our registration model, we conducted experiments using both mean squared error (MSE) and normalized cross-correlation (CC) as loss functions in our filter-based GuidedMorph (Trans). The results, as shown in Table \ref{tab:ablation_table3}, indicate that the choice between MSE and CC does not significantly affect the final outcome with p-value being 0.963. The experiment result is also consistent with those reported in previous studies \cite{voxelmorph,chen2022transmorph}.

\section{Discussion and conclusions}
In this work, we propose the two-stage registration algorithm GuidedMorph. Unlike conventional single-stage registration algorithms that primarily focus on aligning general structures while overlooking intricate internal details, we introduce an additional network that utilizes dense tissue position information to allow the motion from fine but crucial structures—such as dense tissue—to guide the overall registration tracking. Our work provides three approaches for dense tissue information mask extraction: learning-based dense tissue segmentation, breast segmentation, and Frangi vesselness filter–based mask extraction. 

Dense tissue segmentation-based (Ours + DT) setting offers the best dense tissue Dice coefficient, improving the dense tissue Dice by 33.40\% compared to the single-stage baseline with the same mask input. With the significant improvement in aligning the dense tissue, the model didn't compromise with the overall breast Dice, demonstrate with an averaged improvement of 1.4\% in breast Dice coefficient, and 0.33\% in overall breast SSIM on two datasets. 

Utilizing the breast segmentation mask as the second-stage input further improves performance, with a 4.85\% increase in breast Dice coefficient and a 3.90\% improvement in overall breast SSIM, compared to the single-stage baseline using the breast segmentation mask as auxiliary information.

The Frangi vesselness filter–based mask extraction strategy, similar to breast segmentation in providing a rough dense tissue extraction, demonstrates comparable performance in dense tissue Dice when using the breast mask as the second-stage input, and similar breast Dice performance when using the dense tissue mask as additional input. However, the Frangi vesselness filter–based strategy offers our model an option to achieve decent performance without relying on any additional segmentation model. It also provides a relatively fair comparison with previous unsupervised registration models, as neither requires additional model training or inference for segmentation. Experiments in the main result section demonstrate ours (Filter) achieves an improvement of 12.2\% in dense tissue Dice, 5.51\% in breast Dice, and 1.38\% in SSIM compared to the best deep learning–based baseline without relying on external segmentation model input. 

To further eliminate the influence of additional information provided by the filter and to purely evaluate the effectiveness of our two-stage framework, we also compare our (Filter) model with TransMorph under filter setting using the same loss function, hyperparameters, and mask warping algorithm. This comparison is demonstrated in the "First Stage Only/ TransMorph (Filter)" experiment in the ablation study section (see Sec. \ref{sec:Ablation Study}), which shows that the additional stage alone in our framework improves dense tissue Dice by 7.23\%.

Our framework has also been tested with VoxelMorph as the backbone, showing an improvement of 13.01\% in dense tissue Dice, 4.91\% in breast Dice, and 1.21\% in breast SSIM compared to single-stage VoxelMorph, demonstrating the framework’s versatility.

Apart from the two-stage framework, we evaluated two image warping algorithms—applying STN twice and DSTN—and demonstrated that DSTN improves performance by 2.86\% compared to STN twice. We also introduced EDT prior to warping the binary mask during inference to preserve the mask's fine structural details during deformation (see Sec. \ref{sec:methods:distance transformation}). The EDT-based warping algorithm achieved a 2.4\% improvement compared to conventional mask warping.

In the broader field of medical image registration, GuidedMorph’s two-stage framework offers a promising solution for 3D image registration involving large deformations—scenarios in which conventional learning-based methods often fall short \cite{liu2024regfsc, eppenhof2019progressively}. Furthermore, GuidedMorph demonstrates particular effectiveness in breast MRI registration, achieving state-of-the-art performance compared to both learning-based and traditional optimization-based algorithms. This advancement holds significant clinical relevance, as accurate breast registration is essential for longitudinal monitoring \cite{tong2024longitudinal}, treatment response evaluation \cite{rankin2000mri}, and multi-sequence data integration in breast cancer care \cite{nayak2022breast}.

\section*{Acknowledgments}
Research reported in this publication was supported by the National Institute Of Biomedical Imaging and Bioengineering of the National Institutes of Health under Award Number R01EB031575. The content is solely the responsibility of the authors and does not necessarily represent the official views of the National Institutes of Health.

\clearpage
\section*{References}
\bibliographystyle{IEEEtran}
\bibliography{references}
\end{document}